\documentclass[twocolumn,showpacs,preprintnumbers,amsmath,amssymb,
floatfix,nofootinbib]{revtex4-1}
\usepackage{graphicx}
\usepackage{dcolumn}
\usepackage{bm}
\usepackage{amsthm}
\usepackage{amsmath}
\usepackage{amsfonts}
\usepackage{color}

\newcommand{\be}{\begin{equation}}
\newcommand{\ee}{\end{equation}}
\newcommand{\ba}{\begin{eqnarray}}
\newcommand{\ea}{\end{eqnarray}}
\newcommand{\non}{\nonumber}
\newcommand{\n}[1]{\label{#1}}
\newcommand{\eq}[1]{(\ref{#1})}
\newcommand{\pa}{\partial}

\newcommand{\hh}{\, ,\hspace{0.5cm}}
\newcommand{\hhh}{\, ,\hspace{0.2cm}}

\begin{document}
 
\title{Motion of charged particles near weakly magnetized 
Schwarzschild black hole}
\author{Valeri P. Frolov}
\email{vfrolov@ualberta.ca}
\author{Andrey A. Shoom}
\email{ashoom@ualberta.ca}
\affiliation{Theoretical Physics Institute, University of Alberta, 
Edmonton, AB, Canada,  T6G 2G7}
\date{\today}

\begin{abstract}  
We study motion of a charged particle in the vicinity of a
weakly magnetized Schwarzschild black hole and focus on its bounded
trajectories lying in the black hole equatorial plane. If the Lorentz
force, acting on the particle, is directed outward from the black hole,
there exist two qualitatively different types of trajectories, one is
a curly motion and another one is a trajectory without
curls. We calculated the critical value of the magnetic field for the
transition between these two types. If the magnetic field is
greater than the critical one, for fixed values of the particle energy 
and angular momentum, the bounded trajectory has
curls. The curls appear as a result of the gravitational drift. 
The greater the value of the magnetic field,
the larger is the number of curls. 
We constructed an approximate analytical solution for a bounded
trajectory and found the gravitational drift velocity of its guiding center.  
\end{abstract}

\pacs{04.70.Bw, 04.70.-s, 04.25.-g \hfill  
{\bf Alberta-Thy-10-10}}
\maketitle

\section{INTRODUCTION}

There exist both theoretical and experimental indications that a 
magnetic field must be present in the vicinity of black holes.  A
regular magnetic field can exist near a black hole surrounded by
conducting matter (plasma), e.g., if the black hole has an accretion
disk.  The magnetic field near a stellar mass black hole may contain
a contribution from the original magnetic field of the collapsed
progenitor star. The dynamo mechanism in the plasma of the accretion
disk might generate a regular magnetic field inside the disk.  Such a
field cannot ``cross'' the conducting plasma region and is trapped in
the vicinity of the black hole (see, e.g., discussion in
\cite{Punsly}).

Stellar mass and supermassive black holes often have jets, that is
the collimated fluxes of relativistic plasma. It is generally
believed that the MHD of the plasma in strong magnetic and
gravitational fields of the black holes would allow one to understand
formation and energetics of the black holes jets \cite{Punsly}.
Magnetic field in the vicinity of the black holes might play an
important role in  transfer the energy from the accretion disc to 
jets.  Existence of a regular magnetic field near the black holes is
also required for the proper collimation of the plasma in the jets. 

The simplest mechanism for extraction of the rotation energy from black
holes in the presence of magnetic field was proposed by Blandford
and Znajek \cite{BZ}. This mechanism is discussed in detail in the
nice book \cite{para}.  Interesting results of the numerical
simulations, demonstrating  jet formation by the plasma in strong
magnetic and gravitational fields of the black holes, are presented 
in \cite{KSKM} (see also references therein). 
There are some observational  evidences of the existence of
magnetic field around back holes and its effect on the dynamics
of their accretion disks (see, e.g., \cite{K,M}). 

In this paper, we study motion of charged particles in magnetized
black holes. To make our presentation more concrete, we shall use the
estimations for the magnetic field given in \cite{MagneticField}.
Namely, the characteristic scales of the magnetic field $B$ are of
the order of $B_{1}\sim 10^8$G near horizon of a stellar mass
($M_1\sim 10 M_{\odot}$) black hole, and of the order of $B_{2}\sim
10^4$G  near horizon of a supermassive ($M_2\sim 10^9 M_{\odot}$)
black hole.  The value of $B_2$ is also discussed in connection with
the Blandford-Znajek mechanism (see, e.g., \cite{BZ,para}). It should
be emphasized that the aim of the present paper is not to discuss 
special astrophysical objects, but to study interesting features of
the dynamics of a charged particle motion  in the presence of both
the gravitational field of such black holes and the magnetic field in
their vicinity. We use the estimates above to describe only the
domain of the physical parameters, which characterize our system, and
to define the validity of the adopted approximations. 

Let us notice that both the fields $B_1$ and $B_2$  are
weak in the following sense: The space-time local curvature created by the
magnetic field $B$ is of the order of $GB^2/c^4$. It is comparable
to the space-time curvature near a black hole of mass $M$ only if 
\be
\frac{GB^2}{c^4}\sim \frac{1}{r^2_g}\sim \frac{c^4}{G^2 M^2}\, .
\ee
For a black hole of mass $M$ this condition holds if 
\be
B\sim B_M={c^4\over G^{3/2} 
M_{\odot}}\left(\frac{M_{\odot}}{M}\right)\sim
10^{19}(M_{\odot}/M)\mbox{G}\, .
\ee
Evidently, the quantities $B_{1,2}$ for both the stellar mass and
supermassive black holes presented above are much smaller than the
corresponding $B_M$. This means that for our problem the field $B$
can be considered as a test field in the given gravitational
background. Such a magnetic field practically does not
affect motion of neutral particles. 

However, for charged particles, the acceleration induced by the
Lorentz force can be large. The acceleration is of the order of
$qB/(mc)$. Thus, the `weakness' of the magnetic field $B$ is
compensated by the large value of the charge-to-mass ratio $q/m$,
which, for example for electrons, is $e/m_e\approx
5.2728\times10^{17}\,\text{g}^{-1/2}\text{cm}^{3/2}\text{s}^{-1}$. 

In this paper, we consider a Schwarzschild black hole immersed into
magnetic field, which is homogeneous at the spatial infinity, where
it is directed along the `vertical' z-axis.  Close to the black hole
this field is strongly affected by its gravity. Such a model is a
good approximation for more realistic magnetic fields generated by
currents in conducting accretion disk, provided the size of the black
hole is much smaller than the size of the disk (see, e.g., discussion
in \cite{Petterson}).  In our analysis, we shall neglect mutual
interaction of charged particles  moving around the black hole, i.e.,
we consider the approximation of a diluted accretion disk. We also
restrict ourselves to the motion in the equatorial plane of the black
hole, which is orthogonal to the direction of the magnetic field.
Similar approximations were considered in the papers \cite{GP,AG}.

In the presence of the external magnetic field the Lorentz force,
acting on a charged particle, depends on the direction of the
particle motion.  Namely, the presence of the magnetic field breaks
the discrete reflection symmetry in the azimuthal direction, so that motion in
clockwise and counter-clockwise directions is not equivalent.   Many
important results on a charged particle motion near a magnetized
black hole can be found in the interesting review by Aliev and
Gal'tsov \cite{AG}. A study of the marginally stable circular orbits
around a magnetized Kerr black hole can be found in \cite{AO}. 
Motion of charged particles in the dipolar magnetic field around a
static black hole was studied in \cite{Pras1,Preti,Ba}, and around a
rotating black hole in \cite{Pras2}.

To compare the relative strength of the magnetic and gravitational
forces acting on a charged particle moving around the black hole let
us make the following simple estimations: In a flat space-time  (in
the absence of gravity) a particle with charge $q$ and the rest mass
mass $m$ in the magnetic field $B$ has the characteristic  cyclotron
frequency 
\be\n{i1}
\Omega_c=\frac{|qB|c}{E}\,,
\ee
where $E=\gamma\,mc^2$ is energy of the particle. 
Let us compare this frequency with the Keplerian frequency of a
particle orbiting a black hole of mass $M$,
\be\n{i2}
\Omega_K=\frac{r_g^{1/2}c}{r^{3/2}\sqrt{2}}\,.
\ee
Here $r_g=2MG/c^2$ is the black hole gravitational radius.
For $r\sim r_g$ and $\gamma\sim 1$ the ratio of 
these frequencies $\Omega_c/\Omega_K$ is of the order of 
\be\n{kappa}
b\equiv\frac{qBMG}{mc^4}\,.
\ee One can use this dimensionless quantity to characterize the
relative strength of the magnetic and gravitational forces acting on
a charged particle moving near the black hole. For the motion far from the
black hole, the ratio $\Omega_c/\Omega_K$ contains the additional
factor $(r/r_g)^{3/2}$.

For a proton of the mass $m_p$ and the charge $e$ moving near a
magnetized stellar mass or supermassive black hole we have
\ba
b^p_1&=&\frac{eB_1 G M_1}{m_p c^4}\approx 4.7180\times10^{7}\,
,\n{kappa1}\\
b^p_2&=&\frac{B_2 M_2}{B_1 M_1}b^p_1=10^{4}b^p_1\,,\n{kappa2}
\ea
respectively. Both the quantities are large. For electrons their 
value  is larger by the factor $m_p/m_e\approx 1836$. This indicates
that the effect of the magnetic field on a charged particle motion is
not weak.  In a general case, such a  magnetic field essentially
modifies the motion of charged particles. For example, due to this
effect the innermost stable circular orbits around a Schwarzschild black
hole for proper orientation of the particle rotation are shifted
towards its horizon (see, e.g., \cite{GP,AG,AO,RZ}). 

The aim of this paper is to study the dynamics of a charged particle
moving near magnetized black holes. In particular, we shall
demonstrate that in the presence of the magnetic field, there exist
two qualitatively different types of the bounded orbits. In the first
type, the Keplerian motion is modulated by the cyclotron revolution,
so that the particle trajectory has a curly-type structure at small
scales. In the second type the particle trajectory has no curls.  We study the
transition between these two types and calculate the critical value
of the corresponding magnetic field. A combined action of gravity and
electromagnetism  shifts the Keplerian radius and  generates the
drift of the guiding center of the particle trajectory.

This paper is organized as follows: In Sec. II we present equations
of motion of a charged particle moving around a weakly magnetized
Schwarzschild  black hole. In Sec. III we study motion of a charged
particle in a uniform magnetic and weak  gravitational fields.
Section IV contains an analysis of the effective potential
corresponding to a charged particle moving in the equatorial plane of
the black hole. In Sec. V we study the innermost stable circular
orbits. Section VI contains an analysis of bounded trajectories. In
Sec. VII we present an approximate solution for bounded motion of a charged
particle. We summarize our results in Section VIII.  In this paper we
use the sign conventions adopted in \cite{MTW} and units where
$c=1$.      

\section{Equations of motion}

\subsection{Magnetized black hole}  

We shall study a charged particle motion in the vicinity of a
Schwarzschild black hole of mass $M$ in the presence of an
axisymmetric and uniform at the spatial infinity magnetic field $B$.
The Schwarzschild metric reads
\be\n{5}
ds^2=-fdt^2+f^{-1}dr^2+r^2 d\omega^2
\hh
f=1-\frac{r_g}{r}\, ,
\ee  
where $r_g=2GM$ and $d\omega^2=d\theta^2+\sin^2\theta d\phi^2$. The
commuting Killing vectors ${\bm \xi}_{(t)}=\pa/\pa t$ and
${\bm \xi}_{(\phi)}=\pa/\pa \phi$   generate time translations and
rotations around the symmetry axis, respectively. 

Because the metric \eq{5} is Ricci flat, the Killing vectors obey the equation
\be\n{6}
\xi^{a;b}_{\,\,\,\,\,\,\,;b}=0\, .
\ee
This equation coincides with the Maxwell equation for a 
4-potential $A^a$ in the Lorentz gauge
$A^a_{\,\,\,;a}=0$.  The special choice
\be\n{A}
A^{a}={B\over 2}\xi^{a}_{(\phi)}\, ,
\ee corresponds to a test magnetic field, which is homogeneous at the spatial
infinity where it has the strength $B$ (see, e.g., \cite{AG,Wald}). 
In what follows, we assume that $B\geq0$. The electric 4-potential
\eq{A} is invariant with respect to the isometries corresponding to
the Killing vectors, i.e., 
\be
({\cal L}_{{\bm \xi}}A)_{a}=A_{a,b}\xi^{b}+A_{b}\xi^{b}_{\,\,,a}=0\,.
\ee
The magnetic field measured in the rest frame is 
\be
B_{a}=-\frac{1}{2}\varepsilon_{abcd}F^{cd}{{\xi}_{(t)}^{b}
\over |{\xi}_{(t)}^{2}|^{1/2}}
\hhh \varepsilon_{0123}=\sqrt{-g}\, .
\ee
Here $F_{ab}=A_{b,a}-A_{a,b}$ is the electromagnetic field tensor.
The magnetic field corresponding to the 4-potential \eq{A} is
\be\n{7a}
B^{a}\partial_{a}=B\left(1-\frac{r_g}{r}\right)^{1/2}
\left[\cos\theta\frac{\partial}{\partial r}-
\frac{\sin\theta}{r}\frac{\partial}{\partial \theta}\right]\,.
\ee
In the cylindrical coordinates $(\rho,z,\phi)$ at the spatial infinity
it is directed along the $z$-axis. The magnetic field \eq{7a}
coincides with the dipole approximation of the magnetic field of a current
loop of the radius $r=a$ around a Schwarzschild black hole \cite{Petterson}. 
One can show that this approximation is 
valid in the region $r_{g}<r\ll a$.
 
\subsection{Dynamical equations}

Dynamical equation for a charged particle motion is 
\be
m\dot{u}^a=qF^a_{\,\,\,b}\,u^b\,,\n{1}
\ee
where $u^{a}$ is the particle 4-velocity, $u^{a}u_{a}=-1$,
$q$ and $m$ are its charge and mass, respectively. 
Here and in what follows, we denote by the over  dot the derivative 
$d/d\tau$ with respect to the proper time $\tau$. For
the motion in the magnetized black hole (\eq{5} and \eq{A}) there
exist two conserved quantities associated with the Killing vectors:
the energy $E>0$ and the generalized angular momentum
$L\in(-\infty,+\infty)$,  
\ba
&&E\equiv -\xi^a_{(t)}P_a=m\,\dot{t}\left(1-\frac{r_g}{r}\right)\,,\n{8}\\
&&L\equiv \xi^a_{(\phi)}P_a=m\,\dot{\phi}r^2\sin^2\theta
+\frac{qB}{2}r^2\sin^2\theta\,.\n{9}
\ea
Here $P_a=m\,u_a+qA_a$ is the generalized 4-momentum of
the particle. 

It is easy to check that the $\theta$-component of Eq. \eq{1} allows
for a solution $\theta=\pi/2$. This is a motion in the equatorial
plane of the black hole, which is orthogonal to the magnetic field. We
restrict ourselves to this type of motion, for which the conserved quantities
\eq{8} and \eq{9} are sufficient for the complete
integrability of the dynamical equations. 

Let us denote
\be\n{nots}
{\cal E}\equiv\frac{E}{m}\hhh{\cal L}\equiv\frac{L}{m}
\hhh{\cal B}\equiv\frac{qB}{2m}\,,
\ee
where ${\cal E}$ and ${\cal L}$ are the specific energy and the
specific angular momentum. The $r$-component of the dynamical
equation \eq{1} for the planar orbits is
\ba\n{12}
\ddot{r}&=&r\left(\frac{{\cal L}}{r^2}
-{\cal B}\right)\left[\frac{{\cal L}}{r^2}\left(1-\frac{3r_{g}}{2r}\right)
+{\cal B}\left(1-\frac{r_{g}}{2r}\right)\right]-\frac{r_g}{2r^2}\,.\non\\
\ea
The first integral of this equation, which follows from Eqs. \eq{8},
\eq{9}, and \eq{12} is given by
\be\n{13}
\dot{r}^2={\cal E}^2-\left(1-\frac{r_g}{r}\right)
\left[1+r^2\left(\frac{{\cal L}}{r^2}-{\cal B}\right)^2\right]\,.
\ee
The equations for the azimuthal angle $\phi$ and the time $t$ are
\be\n{14}
\dot{\phi}=\frac{{\cal L}}{r^2}-{\cal B}\hh
\dot{t}={\cal E}\left(1-\frac{r_g}{r}\right)^{-1}\, .
\ee
Equations \eq{12}--\eq{14} are invariant under the following
transformations:
\be\n{dt}
{\cal B}\to-{\cal B}\hhh {\cal L}\to-{\cal L}\hhh \phi\to -\phi\, .
\ee
Thus, without loss of the generality, one can assume that the charge
$q$ (and hence ${\cal B}$) is positive. For a particle with a 
negative charge it is sufficient to make the transformation \eq{dt}.

The geodesic motion in the Schwarzschild space-time \eq{5} is well
studied (see, e.g., \cite{Chandrabook,FN,MTW}). Here we study how
this motion is modified in the presence of the magnetic field
\eq{7a}. Our problem is reduced to the set of three first-order
differential equations \eq{13} and \eq{14} containing two independent
parameters ${\cal E}$ and ${\cal L}$, which are integrals of motion.
The solution to the problem depends on the parameters: $r_0$,
$\phi_0$, and $t_0$. They define the initial position of the charged
particle. The radial equation \eq{13} can be integrated
independently.  Substituting its solution into Eq. \eq{14} one can
obtain the angular variable $\phi$ and the time $t$ by integration. 

\section{Weak gravitational field}

\subsection{Flat space-time limit}

Before we study the  motion of a particle near the black hole, it is
instructive to consider an approximation of a weak gravitational
field. In the absence of gravity, when $r_g=0$, equations
\eq{13} and \eq{14} are greatly simplified and take the following form:
\ba
\dot{r}^2&=&
 {\cal E}^2-1-r^2\left(\frac{{\cal L}}{r^2}
-{\cal B}\right)^2 \n{14b}\,,\\
\dot{\phi}&=&\frac{{\cal L}}{r^2}-{\cal
B}\hhh \dot{t}={\cal E}\,.\n{14c}
\ea
The factor ${\cal E}$ in a flat space-time coincides with the Lorentz
$\gamma$-factor.
\begin{figure}[htb]
\begin{center}
\ba
&\hspace{0.1cm}\includegraphics[height=3.0cm,width=3.0cm]{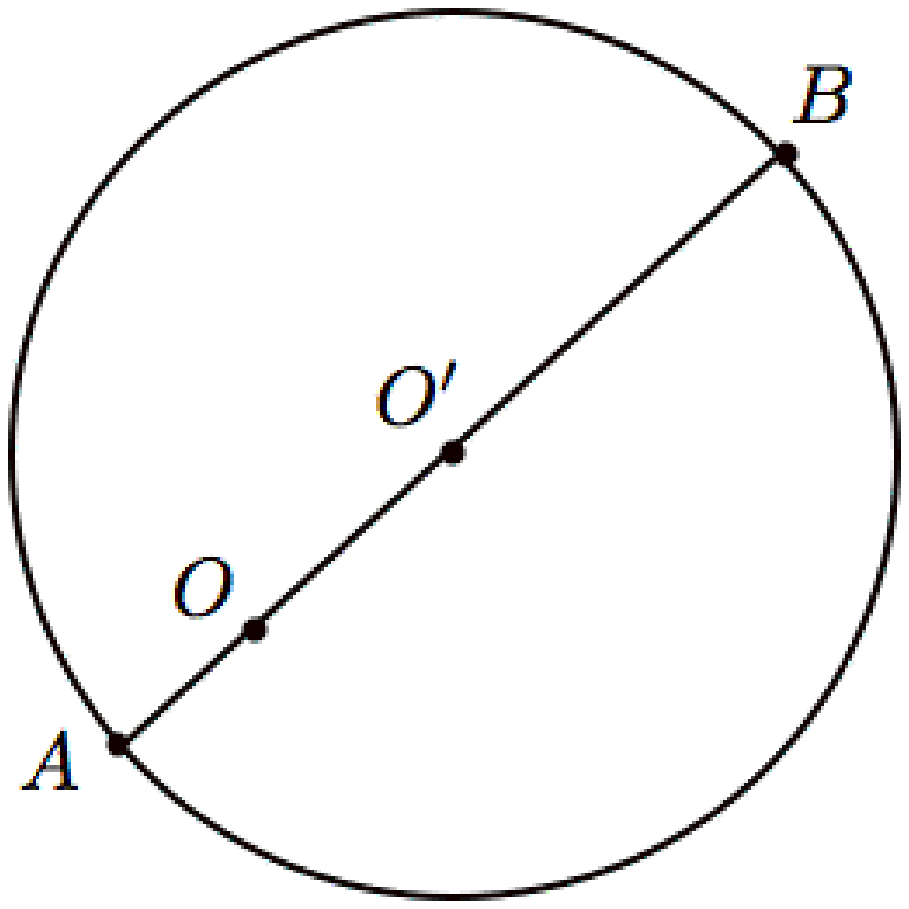}
&\hspace{0.9cm}\includegraphics[height=3.0cm,width=3.5cm]{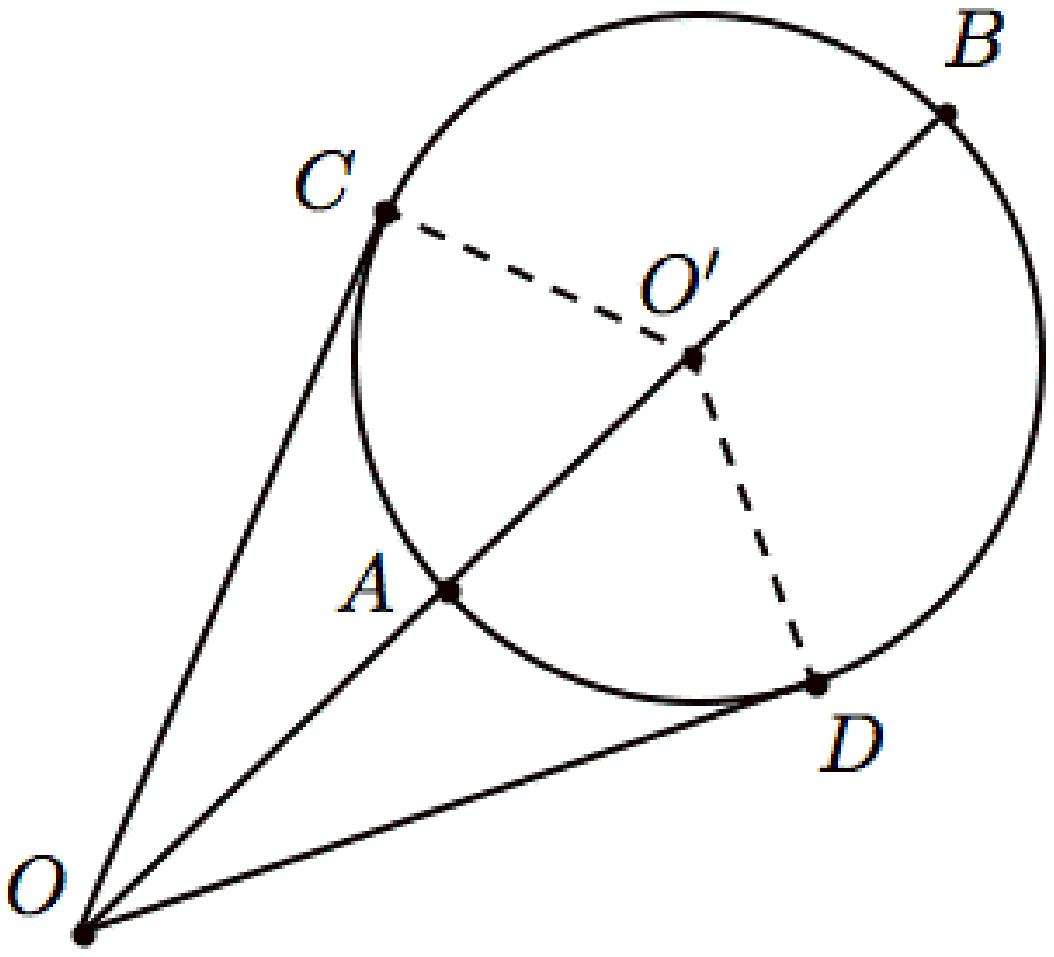}
\non\\
&\hspace{0.1cm}({\bf a}) &\hspace{3cm}({\bf b})\non
\ea
\caption{Motion of a charged particle in a uniform magnetic field
in a flat space-time. Points $A$ and $B$ define the maximal and 
the minimal distances of the particle from the coordinate origin $O$.
({\bf a}): ${\cal L}<0$. ({\bf b}): ${\cal L}>0$. 
Points $C$ and $D$ are the turning
points where $\dot{\phi}=0$.}\label{F0}
\end{center} 
\end{figure} 
The solution to equations \eq{14b} and \eq{14c} is well known. The
particle moves along a circle of the radius 
\be
r_c=\sqrt{|{\cal L}|/{\cal B}}\,,
\ee
where ${\cal L}<0$ is defined with respect to the center of the
circle $O'$.  The corresponding cyclotron frequency \eq{i1} and the
Lorentz $\gamma$-factor are
\be
\Omega_c={2{\cal B}\over {\cal E}}
\hhh {\cal E}=\sqrt{1+4{\cal B}|{\cal L}|}\,.\n{fg}
\ee
The general solution to equations \eq{14b} and \eq{14c}, which
contains two additional constants $r_{0}$ and $\phi_{0}$, can be
easily obtained by coordinate transformation sending the center of
the circle to an arbitrary point $O$ on the plane $z=0$.
Figure~\ref{F0}({\bf a}) illustrates the case when the coordinate
origin is located inside the circle. More interesting for us case is
shown in Fig.~\ref{F0}({\bf b}), where the coordinate origin is
located outside the circle. For such choice of the coordinates the
motion in the $r$-direction is bounded, so that $r_{\text{min}}\le r\le
r_{\text{max}}$, and the angle $\phi$ changes between the two values
$\phi_-$ and $\phi_+$. Motions in the $r$- and $\phi$-direction are
correlated and have the same period, so that the resulting trajectory
is a circle.

\subsection{Approximation of a weak gravitational field}

In the presence of the black hole there is a preferable choice of the
origin $O$ of a coordinate system, namely the black hole location. 
Thus, the symmetry of the flat space-time solution is broken. A black
hole also is a source of the additional (gravitational) force acting
on the particle. Let us now discuss the motion of the particle in the
homogeneous magnetic field, in the approximation when this force is
weak.

The corresponding equations can be obtained from equations \eq{12}
and \eq{14}, assuming that $r_g/r\ll 1$. In the leading order we
have $t\approx{\cal E}\tau$, and 
\ba
\ddot{r}&=&\frac{{\cal L}^{2}}{r^3}- {\cal B}^2 r-g\,,\n{w1}\\
\dot{\phi}&=&\frac{{\cal L}}{r^2}-{\cal B}\,.\n{w2}
\ea
Here $g=r_{g}/(2r^2)$ is the Newtonian gravitational force. This force
is directed along the radius toward the black hole, and is orthogonal to the
magnetic field. 

Take a point $(r_0,\phi_0)$, such that $r_{g}\ll r_{0}$, and introduce the local
Cartesian coordinates $(x,y)$ near it, so that $y$ is directed
along the radius vector at the point, and $x$ is directed ``clockwise''. 
For $x\ll r_{0}$ and $y\ll r_{0}$, which corresponds to short time intervals $\Delta\tau$ and
strong magnetic field ${\cal B}$, we have
\be\n{exp}
r\approx r_0+y\hh \phi\approx\phi_0-x/r_0\, .
\ee
Since $g$ contains the small factor
$r_g/r_0$, it is enough to keep the leading term
\be
g_0=\frac{r_{g}}{2r_0^2}\, .
\ee
We substitute  expansion \eq{exp} into equations \eq{w1} and \eq{w2},
and keep the terms
linear in $x$ and $y$. 

In the zero order, these equations give
\be
{\cal L}={\cal B}r_0^2\, .\n{bl}
\ee
The first order terms give the following dynamical equations:
\ba
\ddot{y}&=&-\Omega^2 y-g_0\, ,\n{y1}\\
\dot{x}&=&\Omega y\, .\n{x1}
\ea
Here $\Omega=2{\cal B}=\Omega_{c}{\cal E}$ is the relativistic
cyclotron frequency.  

\begin{figure}[htb]
\begin{center}
\includegraphics[width=7cm]{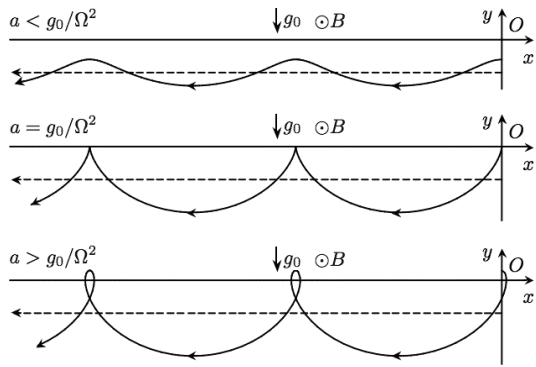}
\caption{Motion of a charged particle in the mutually
orthogonal uniform magnetic and weak gravitational fields. Dashed
line illustrates the motion of the guiding center.}\label{F01} 
\end{center} 
\end{figure} 
The solution to equations \eq{y1} and \eq{x1} corresponding to
the initial conditions $x(0)=0$ and $y(0)=a-g_{0}/\Omega^{2}$ is
\ba\n{ws}
y(\tau)&=&a\cos(\Omega\tau)-{g_0\over \Omega^2}\, ,\\
x(\tau)&=&a\sin(\Omega\tau )-(g_0/\Omega)\tau\, .
\ea
The line on the $(x,y)$ plane described by this solution is called a
trochoid. For $g_0=0$ this solution describes the motion along a
circle with the cyclotron frequency $\Omega_{c}$. In the presence of
gravity the center of the circle moves in the negative $x$-direction
with the velocity
\be\n{v1}
v=r_{0}\langle\dot{\phi}\rangle={g_0\over \Omega}\, .
\ee
The velocity $V$ defined with respect to the rest frame differs by
the Lorentz factor $\dot{t}={\cal E}$,  
\be\n{eV}
V=\frac{v}{{\cal E}}={g_0\over \Omega_c{\cal E}^{2}}\, .
\ee

For $a= g_0/\Omega^2$ the solution takes the form
\ba
y(\tau)&=&{g_0\over \Omega^2}[\cos(\Omega\tau)-1]\, ,\\
x(\tau)&=&{g_0\over \Omega^2}[\sin(\Omega\tau )-\Omega\tau]\, .
\ea
The line on the $(x,y)$ plane described by this solution is called a
cycloid. It separates two different types of motion, with curls (for
$a>g_0/\Omega^2$) and without (for $a<g_0/\Omega^2$) (see
Figure~\ref{F01}). Such a curly-type structure is characteristic for
a charged particle motion in the presence of both magnetic and
non-magnetic forces (see, e.g., \cite{Alfv}, Chapter 2).

The obtained solution has a simple interpretation. In a frame moving
with the velocity \eq{eV} the magnetic field $B$ `generates' the
electric field $E_{y}=\gamma VB$, which is orthogonal to both the
field $B$ and the velocity $V$. This electric field, directed along the
$y$-axis, acts on the
charge $q$ with the force $qE$. The velocity $V$ is defined by the
condition, that this force exactly compensates the gravitational
force $mg_0$. Thus, the motion of a charged particle due to the
gravitational force exerted on it in the direction orthogonal to the
magnetic field is analogous to the motion of the particle in the
constant and mutually orthogonal electric and magnetic fields (see, e.g.,
\cite{Lan}).

\section{Charged particle in the Schwarzschild space-time}

\subsection{Dynamical equations}

We return to our main problem, motion of a charged particle near
a magnetized black hole. It is convenient introduce the following
dimensionless quantities:  
\be
{\cal T}=\frac{t}{r_{g}}\hhh\rho=\frac{r}{r_g}\hhh \sigma=\frac{\tau}{r_g}
\hhh \ell=\frac{{\cal L}}{r_g}\,,\n{18}
\ee
and write ${\cal B}=b/r_g$ (see Eqs. \eq{kappa} and \eq{nots}).
Then, the dynamical equations \eq{13} and \eq{14} take the following
form: 
\ba
&&\hspace{0.8cm}\left(\frac{d\rho}{d\sigma}\right)^2={\cal E}^2-U\,,\n{16}\\
&&\frac{d\phi}{d\sigma}=\frac{\ell}{\rho^2}-b
\hh {d{\cal T}\over d\sigma}=\frac{{\cal E}\rho}{\rho-1}\,.\n{17}
\ea
Here 
\be\n{22}
U=\left(1-\frac{1}{\rho}\right)
\left[1+\frac{(\ell-b\rho^2)^2}{\rho^2}\right]\,.
\ee
is the effective potential. According to the adopted convention, we
have $b\ge 0$.  

The parameter $\ell$ can be positive or negative, $\ell=\pm |\ell|$.
For $\ell>0$ (sign $+$) the Lorentz force, acting on a charged
particle, is repulsive, i.e., it is directed outward from the black
hole, and for $\ell<0$ (sign $-$) the Lorentz force is attractive,
i.e., it is directed toward the black hole\footnote{According to the
terminology used in the papers \cite{GP} and \cite{AG}, ``Larmor"
rotation  corresponds to the Lorentz force pointing toward the black
hole, while ``anti-Larmor" rotation corresponds to the Lorentz force
pointing in the opposite direction.}.

\subsection{Effective potential} 

Let us study properties of the effective potential.  We assume
that the mass of the black hole and the strength of the magnetic
field are fixed. Thus, for a given charged particle the parameter $b$ is also fixed, and the effective
potential is a function of two variables, $\ell$ and $\rho$. We
focus on the motion in the black hole exterior, where $\rho>1$. 
Expression \eq{22} shows that the effective potential is positive in
that region. It vanishes at the black hole horizon, $\rho=1$, and
grows as $b^2\rho^2$ for $\rho\to+\infty$. The latter property implies
that the particle never reaches the spatial infinity, i.e., its motion is
always finite. 

To study the characteristic features of $U$ we use its 
first and second derivatives
\ba
&&\hspace{-0.5cm}U_{,\rho}=\frac{1}{\rho^4}
(2b^{2}\rho^5-b^{2}\rho^{4}-2\ell b\rho^{2}+\rho^{2}-2\ell^{2}\rho+3\ell^{2})\,,\n{ur}\\
&&\hspace{-0.5cm}U_{,\rho\rho}=
\frac{2}{\rho^{5}}(b^{2}\rho^5-\rho^2+2\ell b\,\rho^{2}
+3\ell^{2}\rho-6\ell^{2})\,.\n{urr}
\ea
Extrema of the effective potential are defined by the equation
$U_{,\rho}=0$. The extremal points are the roots of the fifth order
in $\rho$ polynomial in the brackets of expression \eq{ur}. Such 
a polynomial may have five real valued roots. We shall show
now that because of the specific structure of this polynomial, it has
at most two real valued roots in the region
$\rho>1$. 

For this purpose we write the equation $U_{,\rho}=0$ in the following
equivalent form: $P(\rho)=Q(\rho)$, where
\ba
P(\rho)&=&b^{2}\rho^4(2\rho-1)+\rho^2\, ,\\
Q(\rho)&=&2\ell b\,\rho^2+2\ell^2\rho-3\ell^2\, .
\ea
To find a solution to this equation corresponding to the black hole
exterior, it is sufficient to find points of the intersection of the
curves $y=P(\rho)$ and $y=Q(\rho)$ in the interval
$\rho\in[1,\infty)$. Note that in this interval $P(\rho)$ is a positive and monotonically
 growing function, without points of convolution. The second function
represents a parabola, which can either intersect the curve
$y=P(\rho)$ at two points, or do not intersect it at
all. Note that the case when there is only one intersection
in the black hole exterior is excluded for the following reason: In
such a case the potential would have only one extremum (maximum or
minimum), what is impossible for a positive function vanishing at $\rho=0$ and 
growing at infinity. In the limiting case these curves touch each other at one
point, where $P_{,\rho}=Q_{,\rho}$. It is easy to check that at this
point $U_{,\rho\rho}=0$. 

\begin{figure}[htb]
\begin{center}
\includegraphics[width=7cm]{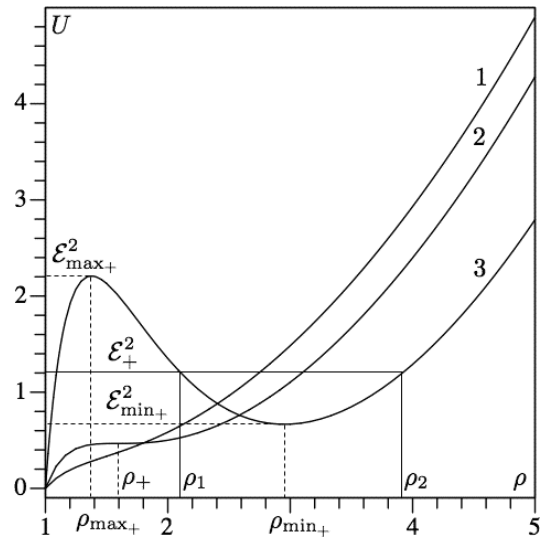}
\caption{Effective potential for $\ell>0$ and $b=1/2$. Curve 1
corresponds to $\ell\approx1.18$. There are no circular orbits. Curve
2 corresponds to $\ell\approx2.07$. There is the innermost stable
circular orbit defined by $\rho=\rho_+\approx1.59$. Curve 3
corresponds to $\ell\approx3.22$. There are unstable and stable
circular orbits corresponding to the minimum and the maximum
of the effective potential. For the bounded trajectory corresponding
to the energy ${\cal E}_{+}$ one has
$\rho\in[\rho_1,\rho_2]$.}\label{F1} 
\end{center} 
\end{figure}

In the former case the intersection of the curves occurs at two points,
$\rho=\rho_{\text{max}}$ and $\rho=\rho_{\text{min}}>\rho_{\text{max}}$. 
The maximum of $U$ is at $\rho=\rho_{\text{max}}$, 
while its minimum is at $\rho=\rho_{\text{min}}$. This means
that at $\rho=\rho_{\text{min}}$ there is a stable circular orbit, while at
$\rho=\rho_{\text{max}}$ there is an unstable one. The condition $U_{,\rho\rho}=0$
singles out the innermost (or marginally) stable circular orbit.
Figure~\ref{F1} illustrates the behavior of the effective potential. 

Let us define the dimensionless parameter
\be\n{alpha}
\omega_{o}^{2}=\frac{1}{2}U_{,\rho\rho}(\rho_{\text{min}})>0\, ,
\ee
which is a measure of the curvature of the effective potential at its
minimum. It is equal to  the square of the frequency of small
oscillations about  a circular orbit in the radial direction.  One
can use the equation $U_{,\rho}=0$ to express $\ell b$ in terms of
other quantities. Substituting this expression into Eqs. \eq{22} and
\eq{alpha}, and eliminating $\ell$ one obtains the following
expression for $\omega_{o}$: 
\be\n{om}
\omega^{2}_{o}=\frac{{\cal E}^{2}_{\text{min}}(\rho_{\text{min}}
-3)}{2\rho^{2}_{\text{min}}(\rho_{\text{min}}-1)^{2}}
+\frac{4b^{2}}{\rho_{\text{min}}}(\rho_{\text{min}}-1)\,,
\ee
where ${\cal E}^{2}_{\text{min}}=U(\rho_{\text{min}})$.
This frequency can be written in the dimensional form as follows: 
\be\n{fr}
\Omega_{o}^{2}=\Omega_{K}^{2}\left(1-\frac{3r_{g}}{r_{\text{min}}}\right)
+\Omega_{c}^{2}\left(1-\frac{r_{g}}{r_{\text{min}}}\right)^{3}\,,
\ee
where the Keplerian frequency $\Omega_{K}$ is defined at
$r=r_{\text{min}}$.   Such a relation was derived in \cite{GP,AG} 
(for a more general case see \cite{Aliev}). It
illustrates a combination of the cyclotron and Keplerian oscillations
due to the gravitational and Lorentz forces acting on a charged
particle. In the Schwarzschild space-time circular orbits
corresponding to $r_{\text{min}}>3r_{g}$ are always stable, i.e.,
$\Omega_{o}^{2}>0$. Marginally stable circular orbits correspond to
$\Omega_{o}=0  $\footnote{The regions of existence and stability of
the circular orbits with respect to small oscillations in the radial
and vertical ($\theta$-) directions where studied in \cite{GP}.}. 

Let us mention another property of the effective potential, which we
shall use later.  We denote
\be
\rho_*\equiv\sqrt{\ell/b}\, ,
\ee
then simple calculations give
\be
U_{,\rho}(\rho_*)=b/\ell\, .
\ee
This means that for bounded orbits and for positive $\ell$ one has
\be\n{rrr}
\rho_*>\rho_{\text{min}}\geq\rho_{\text{max}}\, .
\ee

\begin{figure*}[htb]
\begin{center}
\ba
&&\hspace{-0.5cm}\includegraphics[width=5cm]{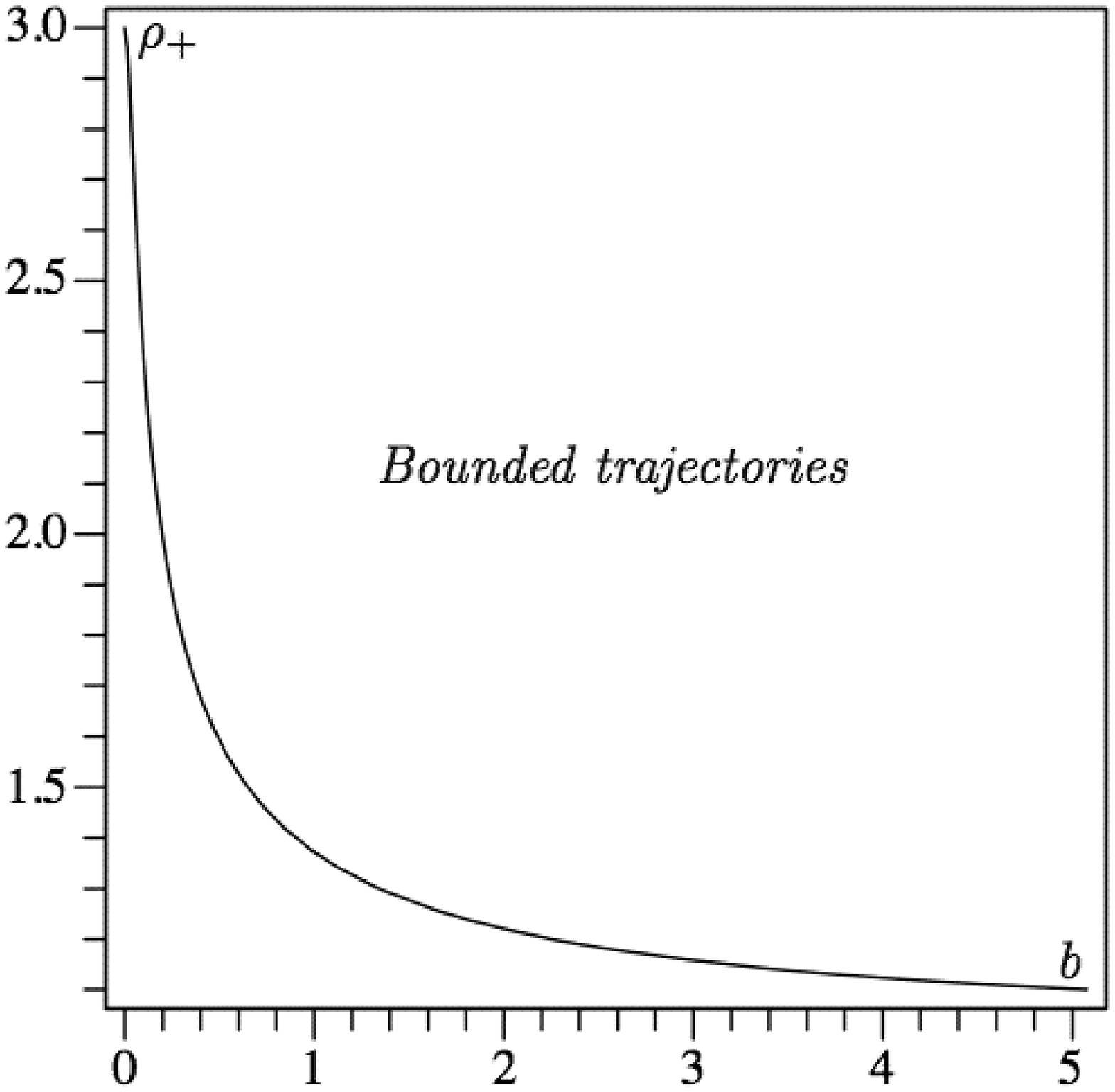}
\hspace{2cm}\includegraphics[width=5cm]{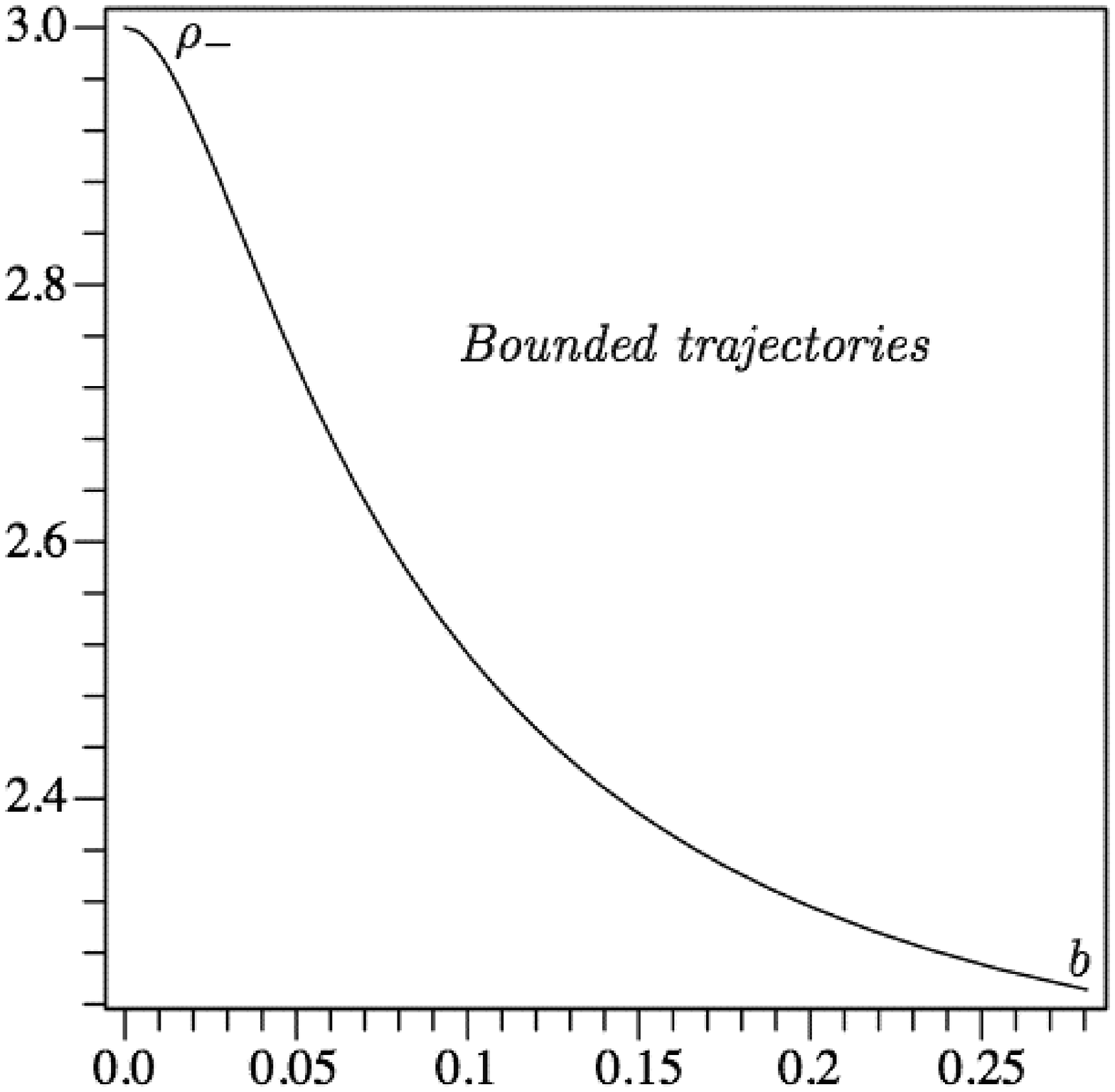}\non\\
&&\hspace{1.95cm}({\bf a})\hspace{6.7cm}({\bf b})\non
\ea
\caption{Dependance of the innermost stable circular orbits
$\rho=\rho_{\pm}$ on $b$.  For each value of $b$ one has
$\rho_+<\rho_-$. For $b\to+\infty$ one has $\rho_{+}\to1$ and
$\rho_{-}\to(5+\sqrt{13})/4\approx2.15$.} \label{F10}
\end{center} 
\end{figure*}
\begin{figure*}[htb]
\begin{center}
\ba
&&\hspace{-0.35cm}\includegraphics[width=4.9cm]{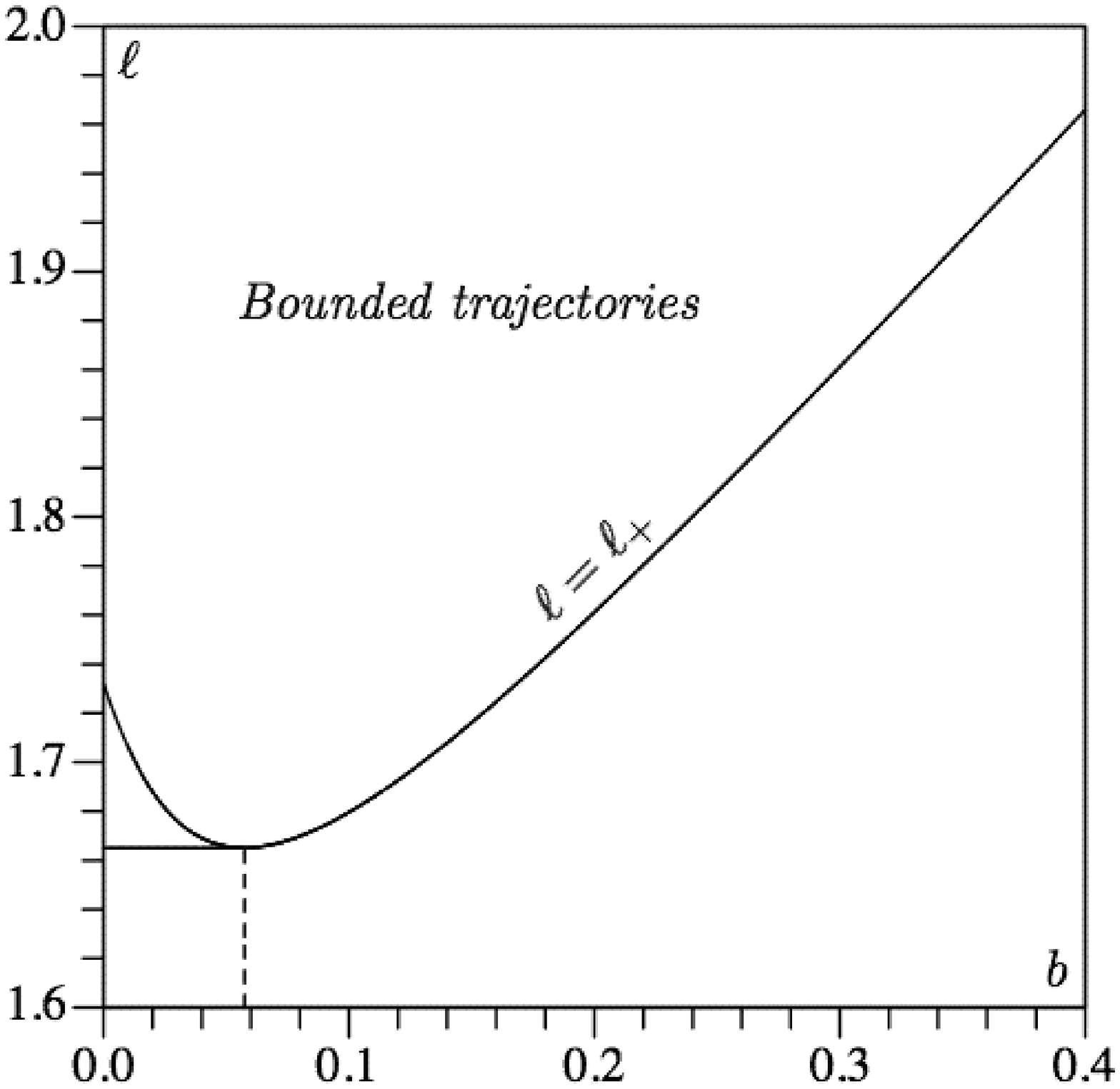}
\hspace{2cm}\includegraphics[width=5cm]{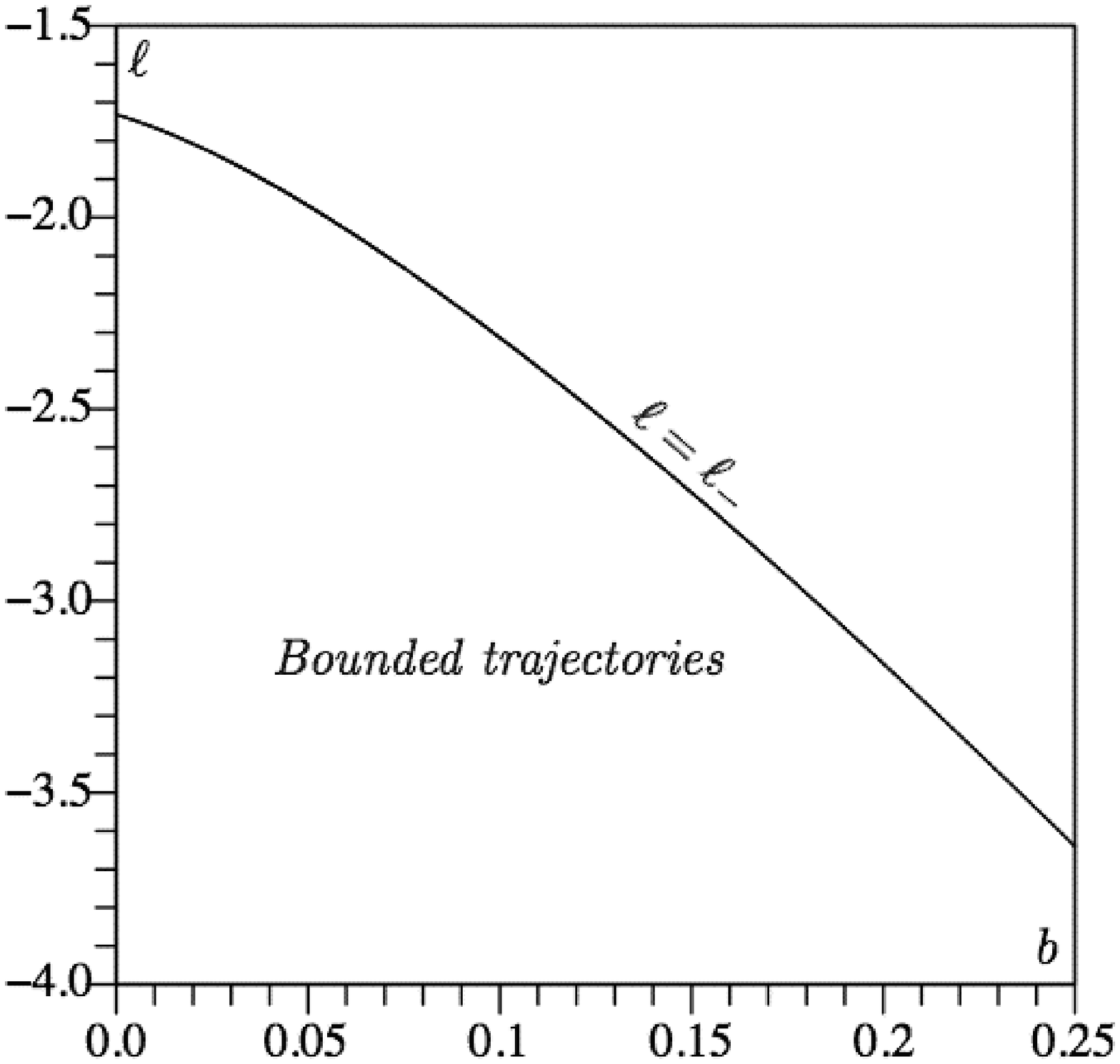}
\non\\
&&\hspace{1.95cm}({\bf a})\hspace{6.7cm}({\bf b})\non
\ea
\caption{Angular momenta $\ell_{\pm}$ for the innermost stable circular
orbits as functions of $b$. ({\bf a}): $\ell>0$, the Lorentz force is
repulsive. ({\bf b}): $\ell<0$, the Lorentz force is attractive; the
curve $\ell=\ell_{-}(b)$ is monotonically decreasing. For $b=0$ we have
$\ell_{\pm}=\pm\sqrt{3}$, what corresponds to a Schwarzschild black
hole. For $b\to+\infty$ we have $\ell_{\pm}\to\pm\infty$.}
\label{F11}
\end{center} 
\end{figure*} 

\section{The innermost stable circular orbits}

The equations $U_{,\rho}=0$ and $U_{,\rho\rho}=0$, which determine the
innermost stable circular orbits, can be presented as follows: 
\ba
&&b^{2}\rho^4(2\rho-1)+\rho^2-2\ell b\rho^{2}-2\ell^{2}\rho+3\ell^{2}=0
\, ,\n{im1}\\
&&b^{2}\rho^5-\rho^2+2\ell b\,\rho^{2}+3\ell^{2}\rho-6\ell^{2}=0\, .\n{im2}
\ea
For a given $b$ these equations allow one to find the parameters
$\rho$ and $\ell$ for such orbits. As usual for this kind of
problems, it is much easier to obtain a solution in an implicit
(parametric) form.  An addition of equations \eq{im1} and \eq{im2}
let us exclude the terms linear in $\ell$. Solving the obtained
equation we derive
\be\n{22c}
\ell_{\pm}=\pm b\,\frac{\rho_{\pm}^2(3\rho_{\pm}-1)^{1/2}}{(3
-\rho_{\pm})^{1/2}}\,.
\ee
The solution exists in the interval $\rho_{\pm}\in(1/3,3]$, where 
$\rho_{\pm}=3$ corresponds to $b=0$. Since we study the black hole
exterior, we have $\rho_{\pm}\in(1,3]$. A substitution of Eq. \eq{22c} into
either of equations \eq{im1} or \eq{im2} gives the following equation
for $\rho_{\pm}$:
\be\n{22d}
4\rho_{\pm}^2-9\rho_{\pm}+3\pm\sqrt{(3\rho_{\pm}-1)(3-\rho_{\pm})}
-\frac{3-\rho_{\pm}}{2b^{2}\rho_{\pm}^2}=0\,.
\ee
This equation allows one to express $b$ in terms of $\rho_{\pm}$,
\be\n{b}
b=\frac{\sqrt{2}(3-\rho_{\pm})^{1/2}}{2\rho_{\pm}\left(4\rho_{\pm}^2
-9\rho_{\pm}+3\pm\sqrt{(3\rho_{\pm}-1)(3-\rho_{\pm})}\right)^{1/2}}\,.
\ee
Substituting this expression into Eq. \eq{22c} we obtain
$l_{\pm}$ as a function of $\rho_{\pm}$,
\be\n{l}
\ell_{\pm}=\pm\frac{\rho_{\pm}(3\rho_{\pm}-1)^{1/2}}{\sqrt{2}
\left(4\rho_{\pm}^2-9\rho_{\pm}+
3\pm\sqrt{(3\rho_{\pm}-1)(3-\rho_{\pm})}\right)^{1/2}}\,.
\ee
The condition that $b$ is real does not impose any new restrictions
on $\rho_+$, so that as above, $1<{\rho}_+\leq3$. For $\rho_-$ it
gives an additional restriction, which is
$(5+\sqrt{13})/4<\rho_-\leq3$. The position of the innermost stable
circular orbit around a Schwarzschild black hole corresponds to
$b=0$, $\rho_{\pm}=3$, and $\ell_{\pm}=\pm\sqrt{3}$. 

Figure~\ref{F10} shows plots of $\rho_{\pm}$ as functions of $b$.
These plots demonstrate that if the magnetic field increases, the
corresponding values of $\rho_+$ and $\rho_-$ decrease.  For a strong
magnetic field and the repulsive Lorentz force, the radius of the
innermost stable orbit may be very close to the gravitational radius
\cite{AO,Preti}. Similar plots for $\ell_{\pm}$ as functions of $b$
are shown in Figure~\ref{F11}. Let us mention that the curve
$\ell=\ell_{+}(b)$ has the minimum
$\ell_{+}=(3+\sqrt{5})^{2}/(8\sqrt{2+\sqrt{5}})\approx1.67$ at
$b=(\sqrt{5}-2)^{3/2}/2\approx0.057$.  

In the limit of the strong magnetic field one has
\ba
\left.\rho_+\right\rvert_{b\gg1}&=&1+\frac{1}{b\sqrt{3}}
+{\cal O}(b^{-2})\,,\n{22g}\\
\left.\rho_-\right\rvert_{b\gg1}&=&\frac{5+\sqrt{13}}{4}+
\frac{41-11\sqrt{13}}{36\sqrt{13}b^2}+{\cal O}(b^{-4})\,,\n{22h}\\
\left.\ell_{+}\right\rvert_{b\gg1}&=&b+\sqrt{3}
+{\cal O}(b^{-1})\,,\n{22g1}\\
\left.\ell_{-} \right\rvert_{b\gg1}&=&-\frac{47
+13\sqrt{13}}{8}b+{\cal O}(b^{-1})\,.\n{22h1}
\ea 
The leading terms in expressions \eq{22g} and \eq{22h} were derived
in \cite{AO,Preti}.  For stable circular orbits one has
$\rho_{\text{min}}>\rho_+$ and $\ell>\ell_+$, if the Lorentz force is
repulsive, and $\rho_{\text{min}}>\rho_-$ and $\ell<\ell_-$, if the Lorentz
force is attractive.

\section{Bounded trajectories}

If the effective potential is a monotonically growing function of
$\rho$, a charged particle always starts its motion in the vicinity
of the black hole, and after reaching the turning point, it falls
down into the black hole. If the effective potential has extrema at
$\rho=\rho_{\text{max,min}}$, then for $\rho<\rho_{\text{max}}$ and ${\cal
E}^2<U(\rho_{\text{max}})$ a charged particle has similar motion. Here we
shall not discuss such type of motion. Instead, we focus on study of
bounded trajectories. For such trajectories the dimensionless radius
$\rho$ changes between its minimal $\rho_{1}$ and maximal $\rho_{2}$
values, $\rho_{\text{max}}\le\rho_{1}\le \rho_{\text{min}}\le \rho_{2}$,  (see
Figure~\ref{F1}). These trajectories  are more interesting for
astrophysical applications, for example for analysis of the motion of
particles in the black hole accretion disk. The specific energy
${\cal E}$ for such trajectories obeys the relation 
\be
{\cal E}_{\text{min}}\leq{\cal E}\leq{\cal E}_{\text{max}}\, .
\ee
The radial motion is periodic with the period
\be
\Delta \sigma_{r}=2\int_{\rho_{1}}^{\rho_{2}} {d\rho\over \sqrt{{\cal
E}^2 -U(\rho)}}\, .
\ee
\begin{figure*}[htb]
\begin{center}
\ba
&\hspace{-1cm}\includegraphics[height=6cm,width=5.1cm]{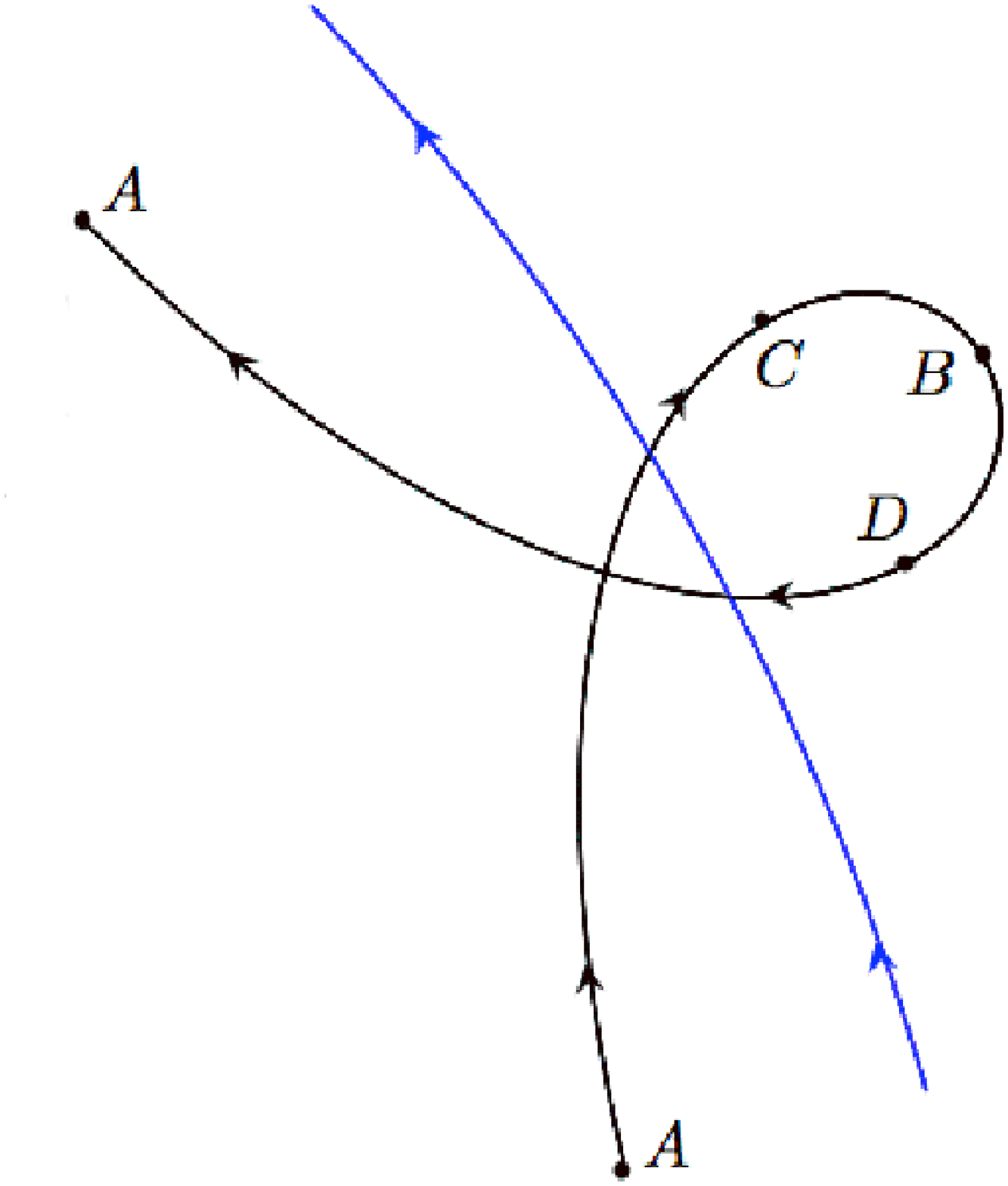}
&\hspace{1cm}\includegraphics[height=6cm,width=4.2cm]{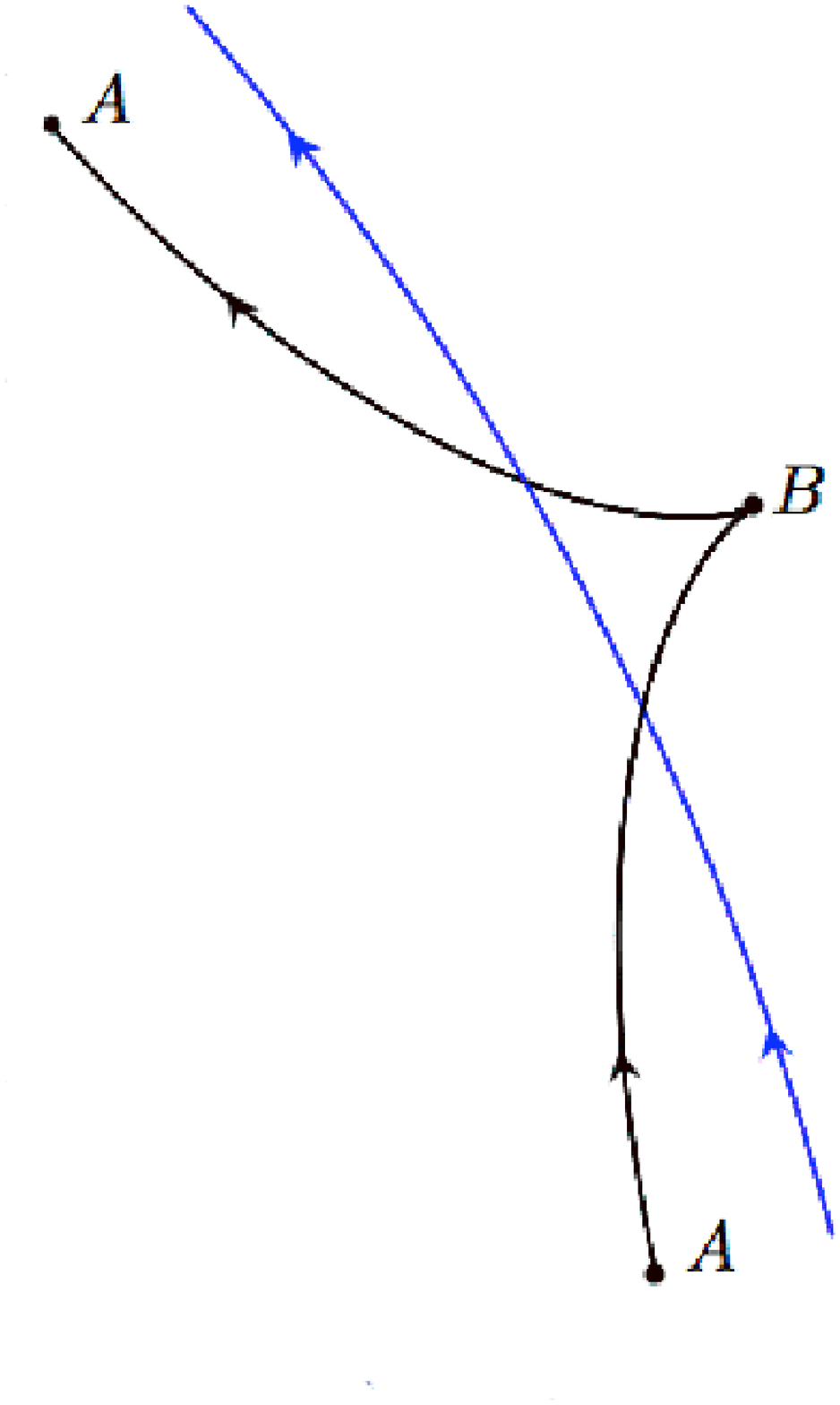}
\hspace{2cm}\includegraphics[height=6cm,width=3.8cm]{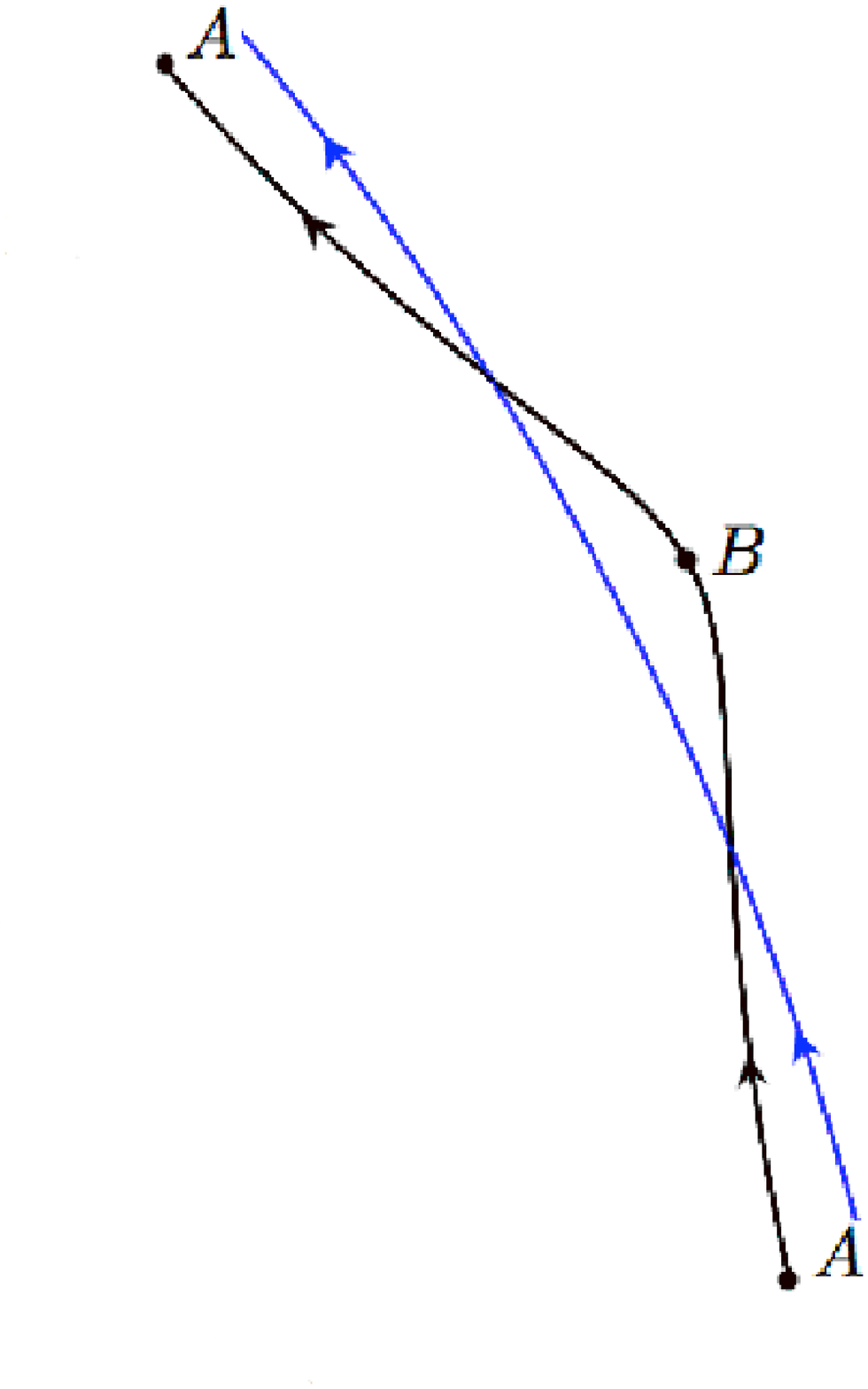}
\non\\
&\hspace{-0.4cm}({\bf a}) &\hspace{3.2cm}({\bf b})\hspace{5.6cm}({\bf c})\non
\ea
\vspace{-0.5cm}
\caption{Types of a bounded trajectory corresponding to $\ell>0$.
Arrows illustrate direction of motion of a charged particle. Circular
arcs represent the stable circular orbit defined by
$\rho=\rho_{\text{min}_+}$. ({\bf a}): ${\cal E}_+>{\cal E}_*$.
Points $A$'s correspond to $\rho=\rho_1$, and point $B$ corresponds to
$\rho=\rho_2>\rho_{*}$. Points $C$ and $D$ are turning points,
where $d\phi/d\sigma=0$. ({\bf b}): ${\cal E}_+={\cal E}_*$. Points
$A$'s correspond to $\rho=\rho_1$, and point $B$ is a turning point,
which corresponds to $\rho=\rho_2=\rho_{*}$, where $d\phi/d\sigma=0$.
({\bf c}): ${\cal E}_+\in[{\cal E}_{\text{min}_{+}},{\cal E}_*)$.
Points $A$'s correspond to $\rho=\rho_1$, and point $B$ corresponds to
$\rho=\rho_2<\rho_{*}$.}\label{F4} 
\end{center} 
\end{figure*} 
\begin{figure}[htb]
\begin{center}
\includegraphics[width=7cm]{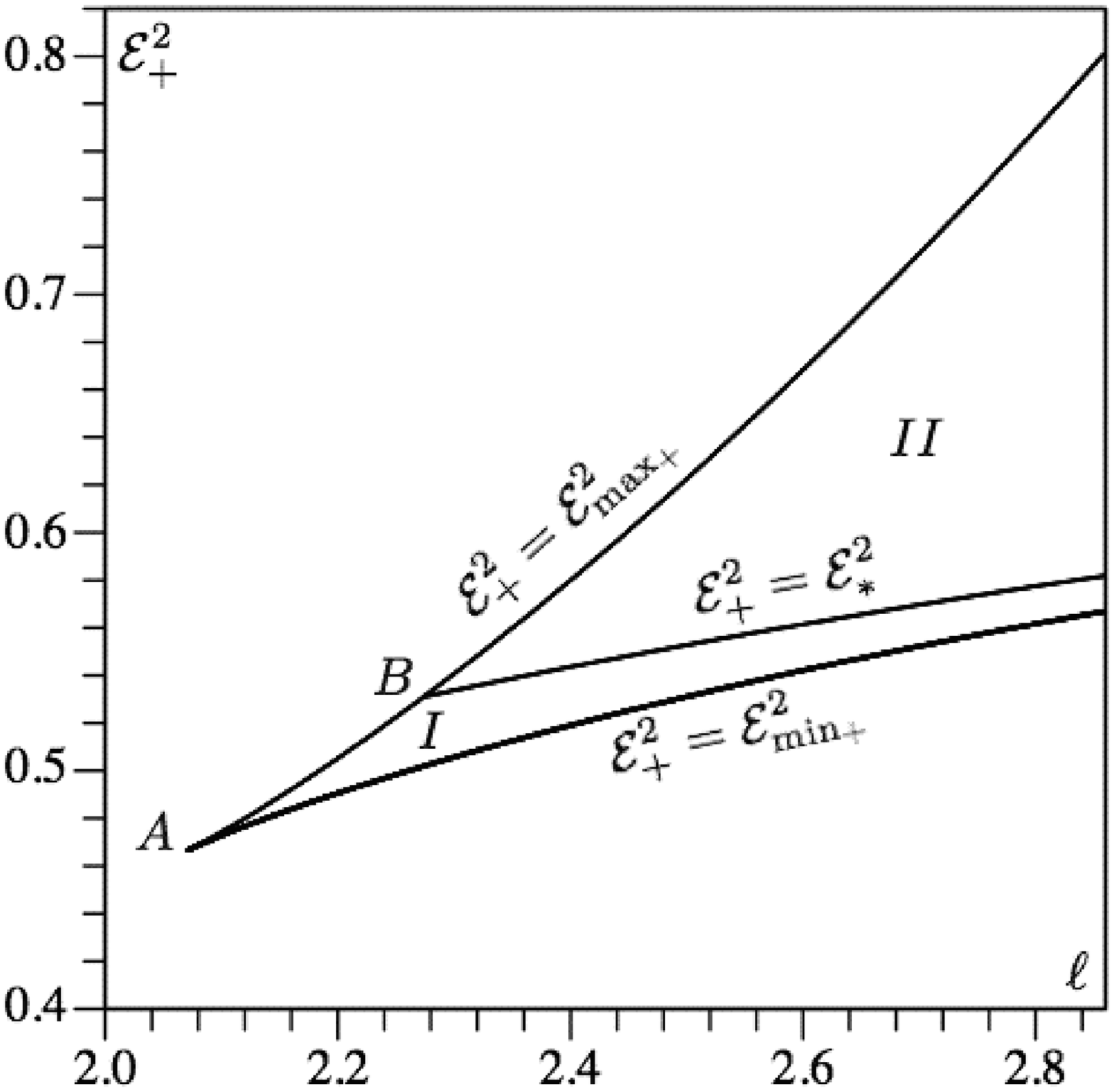} 
\caption{${\cal E}_{\text{max}}^2$, ${\cal E}_{\text{min}}^2$, and ${\cal E}_{*}^2$
as functions of $\ell$ for a fixed value of $b$.  Point $A$, where
the curves ${\cal E}^{2}_{\text{max}}(\ell)$ and ${\cal E}^{2}_{\text{min}}(\ell)$ join,
corresponds to the innermost stable circular orbit. In the domain $I$
between these curves bounded trajectories have no curls. In the  domain $II$
between the curves $ {\cal E}^{2}_{*}$ and ${\cal
E}^{2}_{\text{max}}$ the trajectories are of the curly-type. They
degenerate into unstable circular orbit at point $B$. Bounded motion
occurs in the regions $I$ and $II$. This particular plot is
constructed for $b=1/2$. The qualitative behavior  for different
values of $b$ is the same.}
\label{F55} 
\end{center} 
\end{figure}
Let us consider equation \eq{17} for the angular variable $\phi$
\be\n{feq}
\frac{d\phi}{d\sigma}=\frac{\ell}{\rho^2}-b\,.
\ee
If the Lorentz force is attractive ($\ell<0$) the right hand side of
this equation is always negative. The corresponding motion of a
charged particle is in the clockwise direction.  This motion is
modulated by oscillations in the radial direction. The corresponding
trajectories have no curls.  They are similar  to bounded
trajectories  of test particles moving near a Schwarzschild black
hole (see, e.g., \cite{Chandrabook}). The presence of the magnetic
field just smoothly deforms them. 

From now on we shall focus on more interesting case of the repulsive
Lorentz force ($\ell>0$). In this case there exist two qualitatively
different types of the bounded motion. Let $\rho_1$ and $\rho_2$ be,
as above,  the minimal and the maximal values of $\rho$. If
$\rho_2<\rho_*$ the right hand side of Eq. \eq{feq} is always
positive. The motion has no curls, and $\phi$ monotonically grows with
time. This motion is modulated by oscillations in the radial direction.

For $\rho_2>\rho_*$ the motion of the particle is quite different. 
When the particle moves in the domain $(\rho_1, \rho_*)$, we have
$d\phi/d\sigma>0$, and the angle $\phi$ increases, whereas for the
motion in the domain $(\rho_*, \rho_2)$,  we have $d\phi/d\sigma<0$,
and the angle $\phi$ decreases. The corresponding trajectory has curls.  Increase
of $\phi$, corresponding to $\rho\in[\rho_1, \rho_*)$, is not exactly
compensated by its decrease, corresponding to $\rho\in(\rho_*,
\rho_2]$. As a result, there is a drift of the particle in the
positive $\phi$-direction. The critical type of motion corresponding
to $\rho_2=\rho_*$, for which the trajectory is similar to a cycloid,
is singled out by the condition ${\cal E}={\cal E}_*$, where
\be\n{es}
{\cal E}_*=\sqrt{U(\rho_*)}=\left(1-\frac{1}{\rho_{*}}\right)^{1/2}\, .
\ee
All the three types of bounded trajectories, with curls, the critical,
and without curls, are schematically illustrated in Figure~\ref{F4}.
One can see that the bounded trajectories are  similar to the
trajectories of a charged particle moving in a weak gravitational
field discussed in Section~III. 

For a given $b$  the motion can be specified by the conserved
quantities ${\cal E}$ and $\ell$. Different regions in the $({\cal
E},l)$-plane correspond to different types of motion. Let  us discuss
this in more detail. We assume that $\ell>0$. The condition
$U_{,\rho}=0$ determines $\rho_{\text{max}}$ and $\rho_{\text{min}}$ as functions
of $\ell$.  Substituting these functions into expression \eq{22} for
the effective potential one obtains the corresponding values ${\cal
E}^{2}_{\text{max,min}}=U(\rho_{\text{max,min}},\ell)$. For the limiting value
$\ell=\ell_+$, corresponding to the innermost stable circular orbit,
the two curves ${\cal E}_{\text{max}}(\ell)$ and  ${\cal E}_{\text{min}}(\ell)$
meet each other. For $\ell>\ell_+$, the curve ${\cal E}_{\text{max}}(\ell)$
is always above the curve ${\cal E}_{\text{min}}(\ell)$. At the points of
the extrema of the function ${\cal E}^{2}(\ell)$ one has
\be
{d({\cal E}^2)\over d\ell}=U_{,\ell}={2b\over \rho^2}\left(1-{1\over
\rho}\right)(\rho_*^2-\rho^2)\,,
\ee
where $\rho=\rho_{\text{min,max}}$.
Using Eq. \eq{rrr} we conclude that
\be
{d{\cal E}_{\text{max,min}}^2\over d\ell}>0\, .
\ee
Because $\rho_*$ corresponds to the positive slope of the potential
$U$, one has ${\cal E}_*(\ell)>{\cal E}_{\text{min}}(l)$.  Figure~\ref{F55}
illustrates different types of motion for a given magnetic field
$B$.  Two curves on the `energy-momentum' plane in this plot
correspond to the maximum and the minimum of the effective potential. The
curve ${\cal E}_+^2 ={\cal E}_*^2$ represents the critical energy ${\cal E}_{*}$ as a
function of the angular momentum. The motion for the parameters below
this line, in the domain $I$, is without curls, while the domain $II$
corresponds to trajectories with curls.

Such a plot is convenient, for example for the discussion of the
following problem: Consider a particle with the parameters $E$ and
$L$ in the domain $I$. Let the particle receives an additional
portion of energy $\Delta E>0$ and angular momentum $\Delta L$. This
process moves the point, representing the particle in
$(E,L)$ variables, to a new position. If the point is moved to the
domain $II$, then the excitation changes the trajectory of the
particle, which becomes a curly one.

Using expressions \eq{kappa}, \eq{18}, and \eq{es} we can present the
critical energy ${\cal E}_*$  in the dimensional form as
follows: 
\be\n{cr}
E^{2}_*=m^{2}\left(1-\sqrt{\frac{qBr_{g}^{2}}{2L}}\right)>0\,.
\ee 
This expression establishes a relation between the parameters $E$,
$L$, and $B$, corresponding to the critical motion. For given $B$ and
$L$ the motion is critical if $E=E_*$.  For the critical energy $E_*$
the bounded trajectory has cusps, one of which is illustrated by point $B$ in
Fig.~\ref{F4}({\bf b}).  One can use Eq. \eq{cr} to express the
magnetic field $B$ for the critical motion in terms of the energy $E$
and the angular momentum $L$
\be\n{Bcr}
B_{*}=\frac{2L}{qr_{g}^{2}}\left(1-\frac{E^{2}}{m^{2}}\right)^{2}\,,
\ee
If for fixed values of $E$ and $L$ the magnetic field is larger than
$B_*$, the motion is curly. Thus, keeping the energy and 
angular momentum of a charged particle fixed, one can change the type of its motion by
changing the value of the magnetic field. 
  
\section{Approximate solution for bounded motion}

To analyze properties of bounded trajectories near the minimum of the
effective potential ${\cal E}^{2}_{\text{min}}=U(\rho_{\text{min}})$ one  
can expand $U$ as follows
\be\n{29}
U={\cal E}^{2}_{\text{min}}+\omega_o^2(\rho-\rho_{\text{min}})^2
+...\,,
\ee
where $\omega_{o}$ is defined by Eq. \eq{alpha}, and the dots denote
the omitted higher order terms of the Taylor expansion. In the linear
approximation \eq{29}, equation \eq{16} takes the following form:  
\be\n{eqr}
\left(\frac{d\rho}{d\sigma}\right)^2= {\cal E}^2-{\cal E}_{\text{min}}^2
-\omega_o^2(\rho-\rho_{\text{min}})^2\,.
\ee
Analogous expansion of the angular velocity \eq{17} near
$\rho=\rho_{\text{min}}$ gives the linearized equation for $\phi$
\be\n{eqf}
\frac{d\phi}{d\sigma}=\beta_{0}+\beta_{1}(\rho-\rho_{\text{min}})\,,
\ee
where
\be\n{beta}
\beta_{0}=\frac{\ell}{\rho_{\text{min}}^2}-b\hhh
\beta_{1}=-\frac{2\ell}{\rho^{3}_{\text{min}}}\,. 
\ee
The validity of the linearized approximation requires that
\be\n{lin}
|\rho-\rho_{\text{min}}|\ll \frac{\omega_o^2}{
|U_{,\rho\rho\rho}(\rho_{\text{min}})|}\hhh
|\rho-\rho_{\text{min}}|\ll \rho_{\text{min}}\, .
\ee
We assume that these conditions are satisfied.

Integrating Eqs. \eq{eqr} and \eq{eqf} we derive an approximate
solution for a bounded trajectory of a charged particle 
\ba
\rho(\sigma)&=&\rho_{\text{min}}+A\cos(\omega_o\sigma)\,,\n{32a}\\
\phi(\sigma)&=&\beta_0\sigma +
{\beta_1 A\over \omega_o} \sin(\omega_o\sigma)\,,
\n{32aa}\\
A&=&\frac{\sqrt{{\cal E}^2-{\cal E}_{\text{min}}^2}}{\omega_o}
\,,\n{32b}
\ea
corresponding to the initial conditions
$\rho(0)=\rho_{2}=\rho_{\text{min}}+A$ and $\phi(0)=0$. For $\ell>0$ this solution
describes the motion of a charged particle around the black hole in the
counter-clockwise direction with the average angular
velocity equal to $\beta_0=\ell\rho_{\text{min}}^{-2}-b$. This motion is
modulated by the radial oscillations of the frequency $\omega_o$.
Combination of the radial and azimuthal oscillations with the drift
results in trajectories, qualitatively similar to those shown in
Figs.~\ref{F01} and \ref{F4}. The critical solution is singled out by
the condition that the trajectory has cusps. It happens for
the following amplitude: 
\be
A=-\frac{\beta_0}{\beta_{1}}=\frac{\rho_{\text{min}}}{
2\rho^{2}_{*}}(\rho^{2}_{*}-\rho^{2}_{\text{min}})\, .
\ee
Since $\rho_{*}-\rho_{\text{min}}\ll\rho_{\text{min}}$, this condition implies that
$A\approx\rho_*-\rho_{\text{min}}$ or ${\cal E}\approx{\cal E}_*$, which
corresponds to the exactly calculated bounded trajectory with cusps. 

The average (drift) velocity of the guiding center in the direction of
the increase of $\phi$ is 
\be\n{ev}
v=\rho_{\min} \beta_0={\ell\over \rho_{\text{min}}}-b\rho_{\text{min}}\, .
\ee
The same velocity defined with respect to the time $t$, 
differs from $v$ by the Lorentz factor $\dot{t}$
given in \eq{14}, and is of the form
\be\n{36a}
V=\frac{v}{\dot{t}}=\frac{b(\rho_{\text{min}}-1)}{{\cal E}\rho^{2}_{\text{min}}}
(\rho^{2}_{*}-\rho^{2}_{\text{min}})\,.
\ee

For large values of $b$ (see Eqs. \eq{kappa1} and \eq{kappa2})
equation $U_{,\rho}=0$ can be solved approximately as follows:
\be\n{ap}
\left.\rho_{\text{min}}\right\rvert_{b\gg1}\approx\rho_{*}
-\frac{1}{8b^{2}\rho_{*}(\rho_{*}-1)}\,.
\ee
According to this expression, if $\ell$ is fixed, then $\rho_{\text{min}}$
is a decreasing function of $b$. Thus, strong magnetic field shifts
stable circular orbits toward the black hole horizon.

For $b\gg1$ in the approximation of a weak gravitational field, that is
when $\rho_{\text{min}}\gg1$ and $E\approx mc^{2}$, the drift velocity is 
\be\n{Vw4}
V\approx\frac{1}{4b\rho_{\text{min}}^{2}}=
\frac{mr_{g}}{2qBr_{o}^{2}}=\frac{g_0}{\Omega_{c}}\,.
\ee
where $r_{o}=\rho_{\text{min}}r_{g}$ and $g_0=r_{g}/(2r_0^2)$. This
expression coincides (as expected) with expression \eq{eV} for ${\cal
E}\approx1$.

For $b\gg 1$ in the approximation of a strong gravitational field, that
is $\rho_{\text{min}}\sim 1$, one has 
\be\n{Vs4}
V\propto r_{g}\frac{\Omega^{2}_{K}}{\Omega_{c}}\,,
\ee
where $\Omega^{2}_{K}$ is the Keplerian frequency \eq{i2} calculated
for $r=\rho_{\text{min}}r_{g}\sim r_{g}$. The proportionality
constant in \eq{Vs4} depends on the value of the angular momentum $L$
and is of the order of unity. We see that in both the cases of strong
and weak gravitational fields the gravitational drift velocity is 
proportional to $B^{-1}$. 

Another quantity which characterizes a bounded trajectory corresponding to
the repulsive Lorentz force is the ratio $N$ of the frequency of the
radial oscillations $\omega_{0}$ to  the average angular velocity
$\beta_{o}$.  Using Eqs. \eq{om} and \eq{beta} we obtain
\be\n{38}
N\equiv\frac{\omega_{0}}{\beta_{o}}=\frac{\sqrt{\rho^{4}_{\text{min}}
(3\rho_{\text{min}}-1)+\rho^{4}_{*}(\rho_{\text{min}}-1)}}{\rho^{1/2}_{\text{min}}
(\rho^{2}_{*}-\rho^2_{\text{min}})}\,.
\ee
For $A>-\beta_0/\beta_{1}$ this ratio gives  the number of curls per
one revolution of a particle around the black hole.  This ratio
depends on $\rho_{\text{min}}$ and $\rho_{*}$.  Using the approximation
\eq{ap} for $b\gg1$ we have
\be
\left.N\right\rvert_{b\gg1}\approx8b^{2}\rho^{3/2}_{*}(\rho_{*}-1)^{3/2}\,.
\ee
Thus, the stronger magnetic field, the greater is the ratio and the
number of curls.    

\section{Summary} 

In this paper we studied motion of charged particles in the
equatorial plane of a weakly magnetized Schwarzschild black hole. We
analyzed properties of the corresponding effective potential due to
the combined gravitational and Lorentz forces acting on a charged
particle. We gave a simple analytical proof that this potential
either has two extremal points or none in the black hole exterior. The
critical case, when these extrema coincide, determines a position of
the innermost stable circular orbit (see also \cite{GP,AG,AO}).
We obtained expressions for the radii and the corresponding angular momenta of the
innermost stable orbits in the approximation of the strong magnetic field ($b\gg 1$). 
A similar expression for the radii in the
limit $b\to \infty$ was given before in \cite{AO,Preti}.

In our analysis we mainly focused on the study of bounded trajectories of
charged particles. Such trajectories may be considered as an
approximation to charged particles motion in the black hole accretion
disk, when their mutual interaction is neglected. We constructed an
approximate solution to the dynamical equations which represents a
bounded trajectory localized near the stable circular orbit. 

As a result of our study, we found that if the Lorentz force acting
on a charged particle is repulsive, its bounded trajectory can be
of two different types: with curls and without them. The critical
trajectory, which has cusps, separates these two cases. We calculated
the critical value of the magnetic field for the critical
trajectory. We calculated also the number of curls per one
revolution  and found that its maximal value grows with the increase
of the magnetic field. Using the approximate solution we found the
gravitational drift velocity of the guiding center of a bounded
particle trajectory. 

\begin{acknowledgments}
The authors wish to thank the Natural Sciences and Engineering Research
Council of Canada for the financial support.  One of the authors
(V.F.) is grateful to the Killam Trust for its support. He also
appreciates fruitful discussions of this work at the meeting Peyresq
Physics 15 and is grateful to OLAM, Association pour la Recherche
Fondamentale, Brussels, for its financial support.
\end{acknowledgments}

\end{document}